\documentclass[aps,pra,amssymb,reprint,twocolumn,superscriptaddress,longbibliography]{revtex4-1}
\usepackage[ansinew]{inputenc}
\usepackage{graphicx}
\usepackage[breaklinks=false,colorlinks=true,linkcolor=blue,urlcolor=blue,citecolor=blue]{hyperref}
\usepackage{setspace}
\usepackage[ansinew]{inputenc}
\usepackage{caption}
\usepackage[labelfont=bf]{caption}
\usepackage{graphicx}
\usepackage{amsmath}
\usepackage{color}
\usepackage{lineno}
\usepackage{ulem}
\usepackage{subfigure}
\begin{document}

\title{High-harmonic generation in a strongly overdriven regime} 

\author{B. Major}
\affiliation{ELI-ALPS, ELI-HU Non-Profit Ltd., Wolfgang Sandner utca 3., Szeged 6728, Hungary}
\affiliation{Department of Optics and Quantum Electronics, University of Szeged, D\'om t\'er 9, Szeged 6720, Hungary}
\author{K. Kov\'acs}
\email{kkovacs@itim-cj.ro}
\affiliation{National Institute for Research and Development of Isotopic and Molecular Technologies, Donat str. 67-103, 400293 Cluj-Napoca, Romania}
\author{E. Svirplys}
\affiliation{Max-Born-Institut, Max-Born-Strasse 2A, 12489 Berlin, Germany}
\author{M. Anus}
\affiliation{Max-Born-Institut, Max-Born-Strasse 2A, 12489 Berlin, Germany}
\author{O. Ghafur}
\affiliation{Max-Born-Institut, Max-Born-Strasse 2A, 12489 Berlin, Germany}
\author{K. Varj\'u}
\affiliation{ELI-ALPS, ELI-HU Non-Profit Ltd., Wolfgang Sandner utca 3., Szeged 6728, Hungary}
\affiliation{Department of Optics and Quantum Electronics, University of Szeged, D\'om t\'er 9, Szeged 6720, Hungary}
\author{M. J. J. Vrakking}
\affiliation{Max-Born-Institut, Max-Born-Strasse 2A, 12489 Berlin, Germany}
\author{V. Tosa}
\affiliation{National Institute for Research and Development of Isotopic and Molecular Technologies, Donat str. 67-103, 400293 Cluj-Napoca, Romania}
\author{B. Sch\"utte}
\email{Bernd.Schuette@mbi-berlin.de}
\affiliation{Max-Born-Institut, Max-Born-Strasse 2A, 12489 Berlin, Germany}

\date{\today}

\begin{abstract}
High-harmonic generation (HHG) normally requires a careful adjustment of the driving laser intensity (typically $10^{14} - 10^{15}$~W/cm$^2$) and gas medium parameters to enable good phase matching conditions. In contrast with conventional wisdom, we present experimental results indicating phase-matched HHG in all rare gases, using a high-density medium and a driver laser intensity of around $10^{16}$\,W/cm$^2$. The experimental results are corroborated by theoretical simulations, which indicate that ionization-induced self-phase modulation and plasma defocusing self-regulate the driver laser intensity to a level that is appropriate for good phase matching. A ten-fold broadening of the NIR spectrum is observed, which results in the generation of continuous spectra from $18-140$\,eV in spite of using 50-fs-long driving pulses. The presented scheme represents a simple and versatile concept for the generation of XUV and soft X-ray continua, which are ideally suited for transient absorption and reflection spectroscopy.   
\end{abstract}
\maketitle


In high-harmonic generation, both microscopic and macroscopic effects play an important role. To describe the single-atom response (i.e. the microscopic effects), the semiclassical three-step model has been successfully applied~\cite{corkum93,lewenstein94}. For the macroscopic build-up of HHG, good phase matching conditions are crucial~\cite{constant99,winterfeldt08,popmintchev12}. This requires careful adjustment of a number of experimental parameters of the HHG medium (type of gas, pressure, medium length, density distribution etc.) and the driving laser (intensity, focus position, chirp etc.)~\cite{constant99,heyl16b}, with an intensity that typically lies in the range of 10$^{14}$ - 10$^{15}$\,W/cm$^2$. When generating extreme-ultraviolet (XUV) or soft X-ray pulses at different photon energies, the conditions to achieve optimal phase matching change substantially and typically require different driving laser wavelengths and substantially different gas pressures~\cite{popmintchev12}. 

In addition to the single-atom response and phase matching, reshaping of the driving laser in the HHG medium has obtained increasing attention. Propagation effects have e.g. been exploited for the generation of high-flux XUV pulses in loose-focusing geometries~\cite{tosa03,rivas18,major20}, for the generation of XUV (quasi-)continua~\cite{zeng12, dubrouil14} and for the optimization of HHG in the soft X-ray region~\cite{schutte15e} up to the water window~\cite{johnson18}. Propagation effects were also exploited to induce or enhance an attosecond lighthouse effect~\cite{tosa15, balogh17, tang21} and for the generation of isolated attosecond pulses~\cite{schotz20}. The studies presented in Refs.~\cite{johnson18, schotz20}, in which the driving laser intensity was reduced by a factor up to three during propagation through the HHG medium, were referred to as the overdriven regime.

Here we report on an unusual HHG scheme that may be described as a strongly overdriven regime. We use conditions where the driving laser is focused to intensities around 10$^{16}$\,W/cm$^2$ (in the absence of the HHG medium) in combination with a high-pressure atomic jet that both reshapes the NIR pulses and serves as HHG medium. As a result, a dense plasma is generated which strongly modifies the spatial, spectral and temporal properties of the NIR driving pulses. Our experimental and numerical results show that ionization-induced self-phase modulation (SPM) and plasma defocusing significantly affect the HHG under these conditions. SPM results in a large blueshift and broadening of the driving laser spectrum, while plasma defocusing strongly decreases the NIR intensity at the end of the atomic jet, resulting in phase-matched HHG. This process is self-regulating, because reshaping is more pronounced at higher electron densities. As we demonstrate in this work, the scheme works in all atomic gases and enables the generation of XUV and soft X-ray continua driven by relatively long (50\,fs) NIR pulses. An advantage of this concept is moreover that it allows simple switching between different XUV / soft X-ray spectral regions using different gases, without the need of changing the geometry or time-consuming optimization procedures.

The experiments were performed at the Max-Born-Institut using a setup that is similar to a compact intense XUV setup that we recently demonstrated~\cite{major21}. NIR pulses (central wavelength 790\,nm, pulse energy up to 22\,mJ, pulse duration 50\,fs, repetition rate 10\,Hz) obtained from a Ti:sapphire amplifier~\cite{gademann11} were focused into a gas jet using a spherical lens with a focal length of 1\,m. The $1/e^2$ radius of the NIR pulses before focusing was 10\,mm, and the focused beam waist radius was measured as 42\,$\mu$m. This leads to a focused NIR peak intensity up to $1.6 \times 10^{16}$\,W/cm$^2$ in the absence of a gas medium. The atomic jet was generated by a piezoelectric valve with an orifice diameter of 0.5\,mm that was placed in the NIR focal plane, and HHG was performed directly at the exit of the nozzle. A backing pressure up to 10\,bar was used. The pressure in the interaction region is expected to be lower by a factor of at least three~\cite{drescher18, drescher21}, and the atomic density along the laser propagation direction can be well approximated by a parabolic distribution~\cite{drescher18}. NIR spectra were recorded by placing a screen in the NIR beam path inside vacuum and by recording the scattered light through a vacuum window using an NIR spectrometer. XUV spectra were measured by an XUV spectrometer consisting of a diffraction grating and a microchannel plate / phosphor screen assembly. 

\begin{figure}[tb!]
 \centering
  \includegraphics[width=7.5cm]{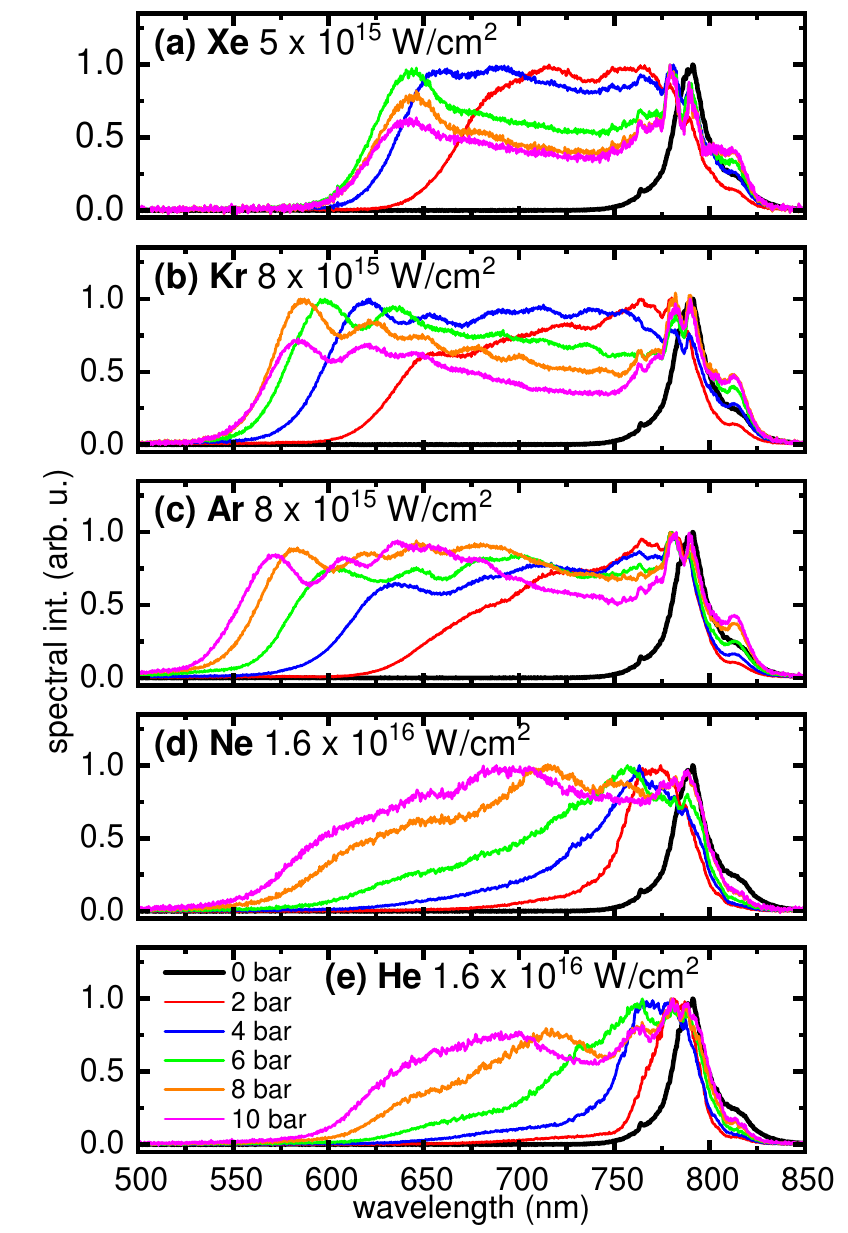}
 \caption{\label{figure_NIR} Spectral broadening of the driving laser pulses for (a) Xe, (b) Kr, (c) Ar, (d) Ne and (e) He using backing pressures from 2\,bar to 10\,bar. The spectra are normalized, and the applied peak intensities are indicated. The black spectrum corresponds to the unperturbed NIR spectrum.}
\end{figure}

\begin{figure}[tb!]
 \centering
  \includegraphics[width=7.5cm]{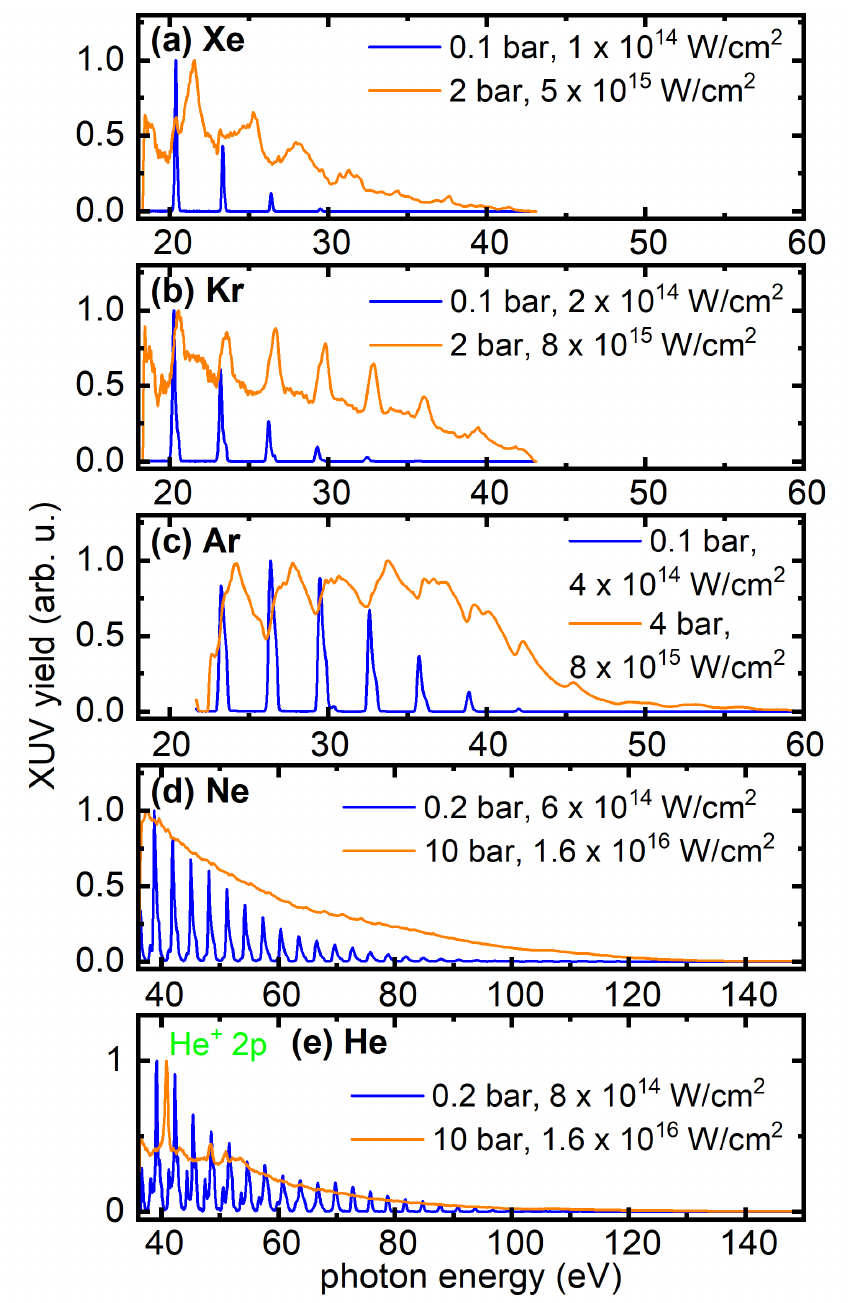}
 \caption{\label{figure_HHG} Normalized HHG spectra obtained from (a) Xe, (b) Kr, (c) Ar, (d) Ne and (e) He. The blue curves represent spectra recorded at standard HHG conditions (i.e. moderate driving laser intensities and gas pressures), resulting in narrowband harmonics. The orange curves represent HHG spectra obtained in the high-pressure jet ($2-10$~bar) at NIR driving laser intensities of $0.5 \times 10^{16}$\,W/cm$^2$ (Xe), $0.8 \times 10^{16}$\,W/cm$^2$ (Kr, Ar) and $1.6 \times 10^{16}$\,W/cm$^2$ (Ne, He). In (e) narrowband lines are visible within the otherwise continuous HHG spectrum (orange curve), which are attributed to NIR-induced free-induction decay~\cite{beaulieu16} involving the $2p$ and higher excited states of He$^+$ ions.}
\end{figure}

\begin{figure*}[htb]
 \subfigure{\includegraphics[width=0.6\textwidth]{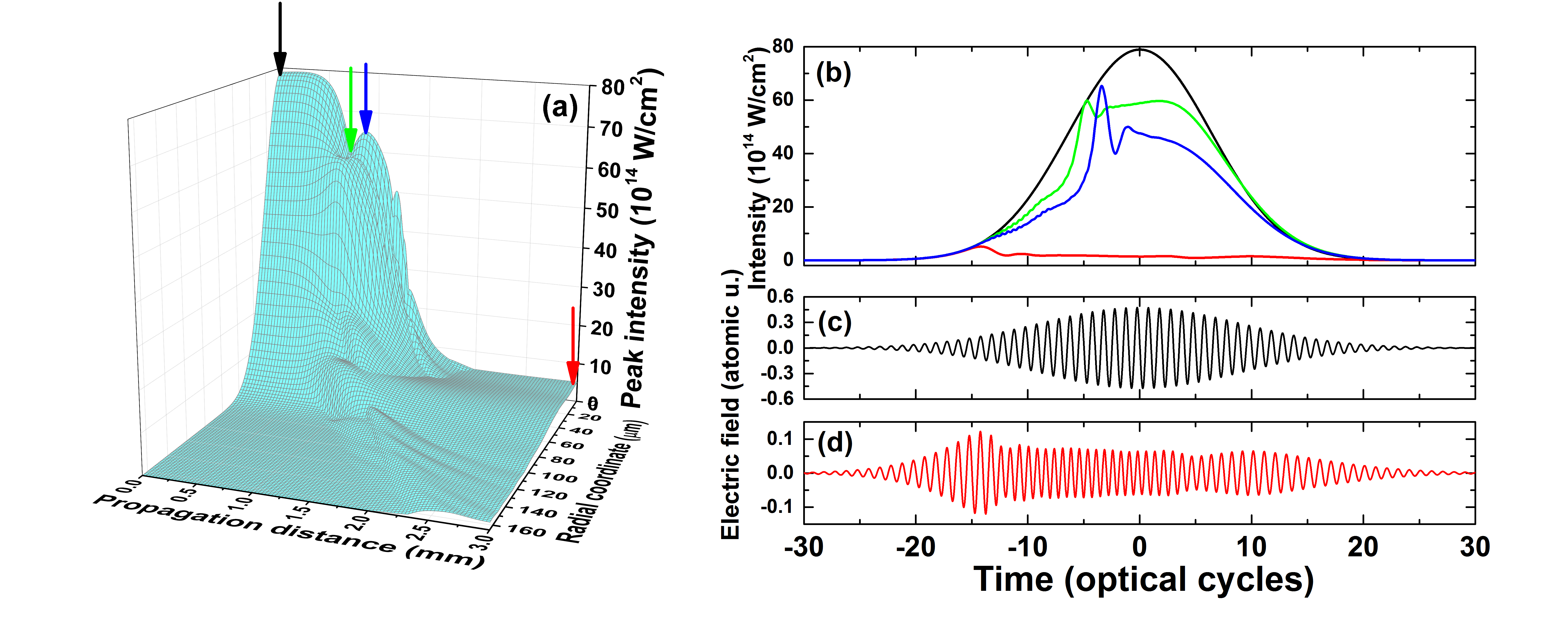}} 
 \subfigure{\includegraphics[width=0.3\textwidth]{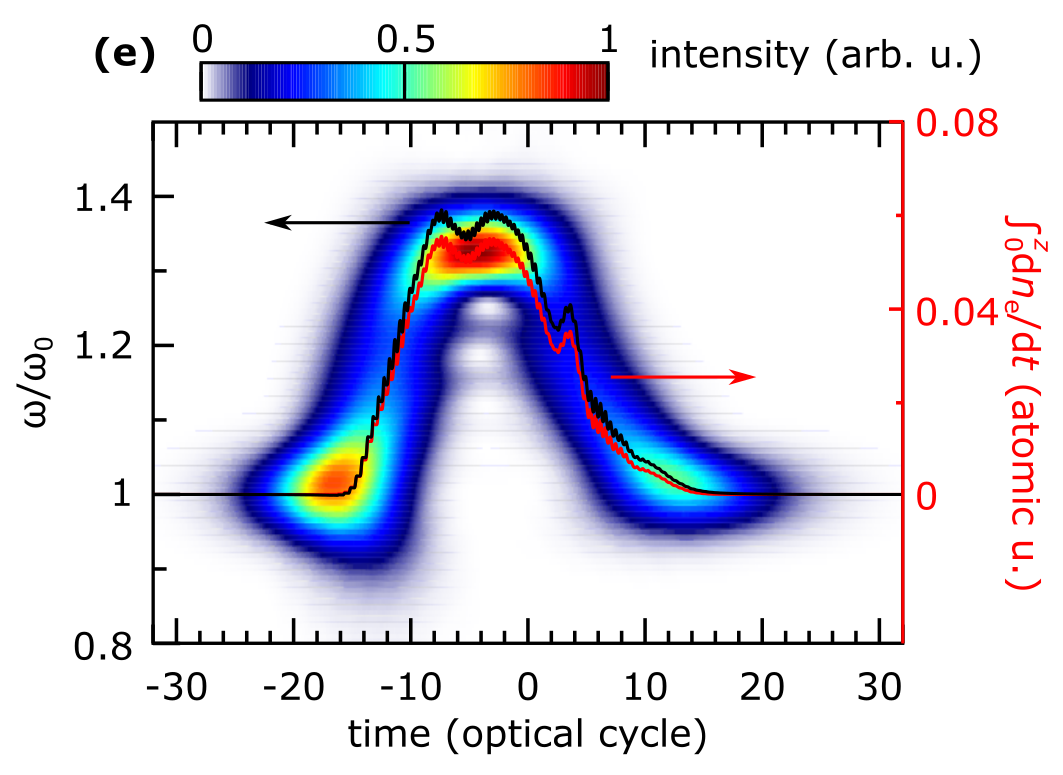}} 
 \caption{\label{figure_sim_NIR1} Simulated properties of NIR pulses with an initial peak intensity of $8 \times 10^{15}$\,W/cm$^2$ during and after propagation through a Ne jet using a peak pressure of 400\,mbar. (a) Radially resolved NIR intensity distribution as a function of the propagation distance $z$ through the jet. The arrows mark specific $z$ values for which the on-axis temporal intensity distributions are shown in (b). The electric field is shown (c) before and (d) after propagation through the jet. (e) Driving laser spectrum as a function of time. The black curve shows the cycle-averaged carrier frequency calculated from the temporal change of the refractive index, which is dominated by changes of the density of free electrons (red curve).}
\end{figure*}

Fig.~\ref{figure_NIR} shows that the NIR driving pulses are substantially broadened and blue-shifted after propagation through a dense jet consisting either of Xe, Kr, Ar, Ne or He. In addition to the atomic species, the only parameter that was changed was the NIR pulse energy. Higher pulse energies were used for atomic species with higher ionization potentials to generate sufficiently dense plasmas which are required for efficient reshaping. The NIR peak intensity ranged from $5 \times 10^{15}$\,W/cm$^2$ for Xe to $1.6 \times 10^{16}$\,W/cm$^2$ for Ne and He. In the investigated parameter regime, the largest broadening is observed in Ar, where the full width at half maximum (FWHM) is increased from 20\,nm to 250\,nm using a backing pressure of 10\,bar. In the other gases the spectral width is increased by one order of magnitude as well. It is evident that atomic species with higher ionization potentials require higher gas pressures and higher intensities to achieve a similar broadening effect. Moreover, we have observed an increased divergence of the NIR pulses in the presence of the atomic jet. For instance, in the case of Ne using a backing pressure of 10\,bar, the divergence of the driving laser was increased by a factor of about four.

Spatially integrated HHG spectra obtained under the conditions of Fig.~\ref{figure_NIR} are displayed in Fig.~\ref{figure_HHG} for Xe, Kr, Ar, Ne and He (orange curves). For comparison, HHG spectra obtained at lower intensities and lower gas densities are shown as well (blue curves), where individual, narrowband harmonics are clearly discernible. It is evident that when going from a standard HHG regime (blue curves) to the strongly overdriven regime (orange curves), the harmonic spectra are substantially broadened. This results in (quasi-)continuous XUV and soft X-ray spectra spanning the range from 18\,eV to 140\,eV when using different atomic species. In those cases where individual harmonics are still visible in the strongly overdriven regime (orange curves in Fig.~\ref{figure_HHG}(a)-(c)), clear spectral blueshifts are observed with respect to the harmonics in the standard HHG regime (blue curves). The XUV pulse energies are estimated as about 150~nJ for HHG in Xe and Kr using an XUV photodiode~\cite{major21}. We note that a comparison of the flux in the two regimes is not meaningful, as HHG in the standard regime was not optimized for flux.

To understand the physics leading to HHG in the strongly overdriven regime, we have performed extensive numerical calculations in Ne using the adapted version of a three-dimensional non-adiabatic model described in Refs.~\cite{tosa05a, tosa05b, major19}. The dense plasma formation made it necessary to restrict the calculations to an NIR intensity of $8 \times 10^{15}$\,W/cm$^2$ and a maximum pressure of 400\,mbar. 

Fig.~\ref{figure_sim_NIR1}(a) shows the radially resolved NIR intensity distribution as a function of the propagation distance $z$ within the atomic jet using a peak pressure of 400\,mbar. The NIR peak intensity decreases rapidly during propagation through the jet due to plasma defocusing and absorption, from an initial value of $8\times 10^{15}$\,W/cm$^2$ (black arrow) to a final value of $5\times 10^{14}$\,W/cm$^2$ (red arrow). At the same time, the beam radius is increased from 45\,$\mu$m to 192\,$\mu$m. We note that for a peak pressure of 133\,mbar, a final NIR intensity of $6\times 10^{14}$\,W/cm$^2$ is reached, indicating that the process is self-regulating. The NIR peak intensity does not decrease monotonically as a function of $z$, but exhibits intermediate maxima and minima, which are exemplary indicated by the blue and green arrows. Temporal intensity distributions at these positions are depicted in Fig.~\ref{figure_sim_NIR1}(b), showing that the average on-axis intensity decreases with increasing propagation distance through the jet. The NIR electric field after propagation through the jet (Fig.~\ref{figure_sim_NIR1}(d)) becomes strongly modified with respect to the initial electric field (Fig.~\ref{figure_sim_NIR1}(c)) starting at around $-14$ optical cycles. The decreased oscillation period is the result of ionization-induced SPM~\cite{bloembergen73, he14} and manifests itself as a spectral blueshift. This is corroborated by Fig.~\ref{figure_sim_NIR1}(e), which compares the time-dependent driving laser spectrum that was extracted from the calculations, to a prediction of the spectral shift as a result of the temporal change of the refractive index (black curve, see Ref.~\cite{penetrante92}). For comparison, changes of the free-electron density obtained from the simulations are shown by the red curve. This contribution dominates the change of the refractive index.

\begin{figure}[htb]
 \centering
  \includegraphics[width=7cm]{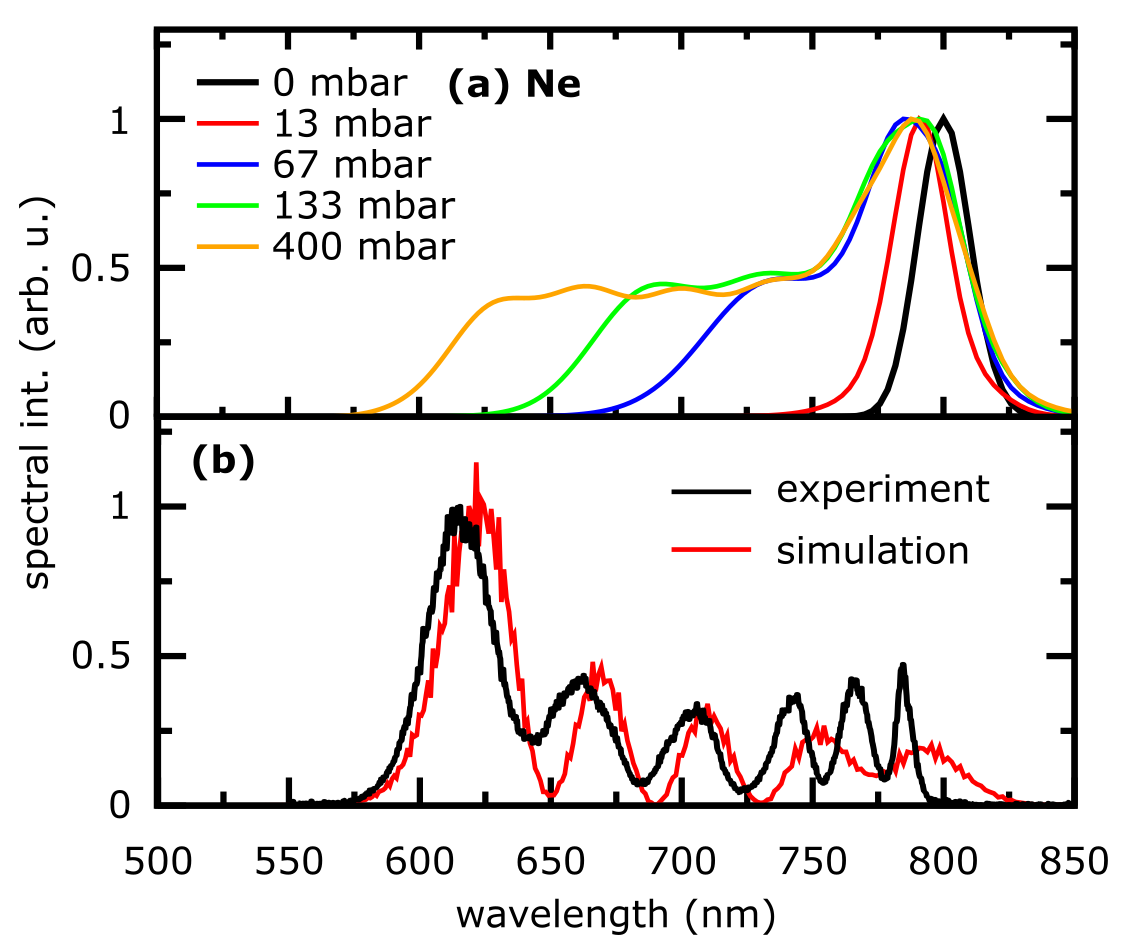}
 \caption{\label{figure_sim_NIR2} (a) Simulation of the spatially integrated NIR spectra after propagation through the jet at different peak pressures. (b) Contribution of the NIR spectrum at an angle of 19\,mrad obtained in the experiment (black curve) and in the simulation (red curve), showing clear spectral modulations in both cases.}
\end{figure}

Spatially integrated NIR spectra after propagation through the jet are presented in Fig.~\ref{figure_sim_NIR2}(a) for different pressures. The obtained blueshift is in good agreement with the experimental results shown in Fig.~1. Note, however, that the NIR spectrum is strongly spatially dependent. When selecting a small part of the driving laser pulse using an iris at an angle of 19~mrad from the optical axis (where no light is present when the gas jet is switched off), the spectrum shows strong modulations both in the experiment (black curve in Fig.~\ref{figure_sim_NIR2}(b)) and in the simulations (red curve). We attribute these modulations to interference resulting from the fact that each frequency component occurs at two different times within the pulse (cf. Fig.~\ref{figure_sim_NIR1}(e))~\cite{shimizu67}. 

\begin{figure}[tb]
 \centering
  \includegraphics[width=7cm]{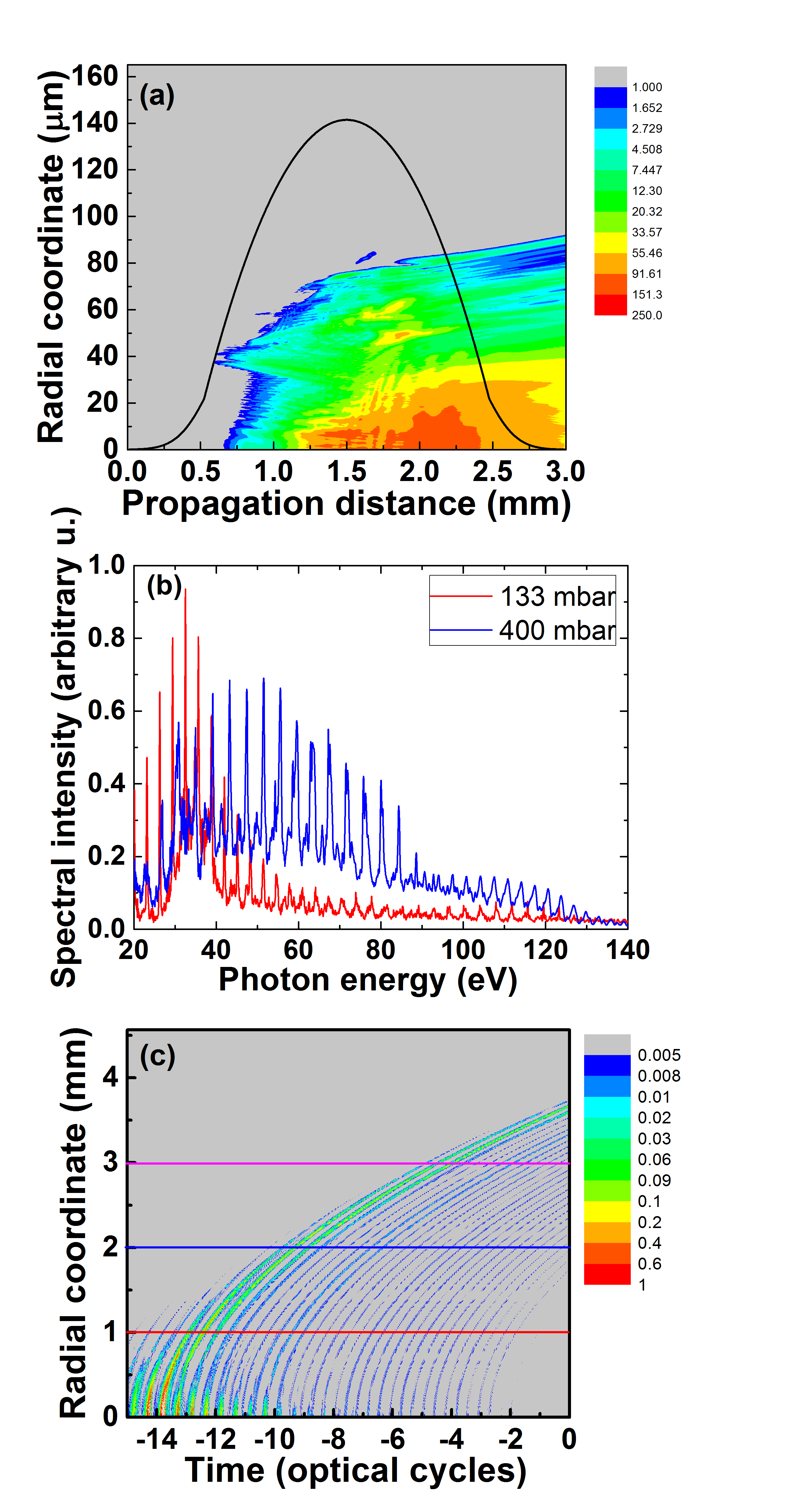}
 \caption{\label{figure_sim_XUV} Simulated HHG properties in the strongly overdriven regime. (a) HHG build-up in the spectral region from $42-45$\,eV as a function of the propagation distance and radial coordinate shown on a logarithmic scale. The black curve shows the parabolic pressure distribution. (b) Spatially integrated HHG spectra at pressures of 133\,mbar (red curve) and 400\,mbar (blue curve), showing the generation of a quasi-continuous XUV spectrum in the latter case. (c) Radially dependent HHG emission for harmonic orders 15-51 as a function of time. The data were analyzed at a distance of 70\,cm from the jet and are shown on a logarithmic scale. The horizontal lines indicate that spatial filtering may result in the generation of short attosecond pulse trains.}
\end{figure}

Simulated properties of the generated harmonics are presented in Fig.~\ref{figure_sim_XUV}. The build-up of the harmonics in the energy range from $42-45$\,eV (Fig.~\ref{figure_sim_XUV}(a)) occurs predominantly at propagation distances between 1 and 2~mm, where the NIR intensity (Fig.~\ref{figure_sim_NIR1}(a)) has already substantially dropped. The XUV spectrum evaluated 70\,cm behind the jet (Fig.~\ref{figure_sim_XUV}(b)) obtained at a pressure of 133\,mbar (red curve) shows narrowband harmonics, whereas the spectrum obtained at 400\,mbar (blue curve) is quasi-continuous, in qualitative agreement with the experimental results. Note that harmonics are still clearly visible in this simulated spectrum (in contrast to the experimental spectrum shown as orange curve in Fig.~\ref{figure_HHG}(d)), which is attributed to the lower pressure that was used in the simulation as compared to the experiment. HHG emission in the far field as a function of time and the radial coordinate is shown in Fig.~\ref{figure_sim_XUV}(c). On one hand, a contribution is visible at radii $<0.7$\,mm which is peaked around $-14$ optical cycles, corresponding to the time at which the NIR intensity is highest (cf.~Fig.~\ref{figure_sim_NIR1}(b)). On the other hand, a contribution appears at larger radii, which around $-13$ optical cycles starts to be clearly separated from the contribution at small radii. In this regime, the divergence of each consecutive attosecond burst increases, which is reminiscent of the attosecond lighthouse effect~\cite{vincenti12,kim13,tosa15}. The bending of the individual attosecond bursts towards larger radii is a result of the wavefront curvature of the XUV beam in the far field. The horizontal lines in Fig.~\ref{figure_sim_XUV}(c) indicate that spatial filtering in the far field could be applied to obtain relatively short attosecond pulse trains.

In this work we have shown that NIR driving laser intensities far above the typical intensities can be used for phase-matched HHG by exploiting substantial spatial and spectral reshaping of the driving laser in a high-pressure atomic jet. Experimental signatures of this reshaping were a large spectral blueshift and a broadening of both the NIR spectra and the generated harmonics, and an increased divergence of the driving laser pulses. These results were reproduced by numerical calculations, which further showed that substantial temporal reshaping takes place in the gas jet. Both experiments and simulations indicated that this is a self-regulating process, where strong reshaping takes place until the plasma density generated by the driving laser becomes sufficiently low to suppress further reshaping, thereby providing good conditions for phase matching in the subsequent HHG. As a result, the scheme works in all atomic gases with only a few simple changes of the experimental parameters.


Our approach represents a simple method for the generation of continuous XUV and soft X-ray spectra by HHG using long driving laser pulses. As such it is ideally suited for transient absorption and reflection spectroscopy in atomic, molecular and solid-state targets~\cite{geneaux19}. Furthermore, our results suggest that it is possible to spatially select a short attosecond pulse train. These few-femtosecond XUV pulses might be used in combination with pump-probe techniques in which the temporal resolution is not limited by the pulse duration of the probe laser. One example is terahertz streaking that can be used to used to resolve few-femtosecond dynamics in spite of the much longer picosecond durations of the terahertz pulses~\cite{schutte12}.

\bibliography{Bibliography}

\end{document}